\documentclass{anstrans}

\usepackage{graphicx, booktabs, algorithm, algorithmicx, algpseudocode, microtype}

\algnewcommand\algorithmicestep{\textbf{E-Step:}}
\algnewcommand\EStep{\State\algorithmicestep}
\algnewcommand\algorithmicmstep{\textbf{M-Step:}}
\algnewcommand\MStep{\State\algorithmicmstep}

\title{Diversion Detection in Partially Observed Nuclear Fuel Cycle Networks}
\author{Elizabeth Hou$^{*}$, Yasin Y{\i}lmaz$^{\dagger}$, Alfred O. Hero$^{\dagger}$}

\institute{
$^{*}$University of Michigan, Department of Statistics, 1085 S. University Ave, Ann Arbor, MI 48109
\and
$^{\dagger}$University of Michigan, Department of EECS, 1301 Beal Ave, Ann Arbor, MI 48109
}

\email{emhou@umich.edu \and yasiny@umich.edu \and hero@eecs.umich.edu}

\begin{document}
\section{Introduction}
A nuclear fuel cycle contains several facilities with different purposes such as mining, conversion, enrichment, and fuel rod fabrication. These facilities form a network, which is naturally sparse in the number of connections (i.e., edges) since not every facility directly interacts with all the others. Given the knowledge of a network baseline, we are interested in detecting anomalous activities in this network, which may signal the diversion of nuclear materials. Anomalies can take the form of a new or missing edge or abnormal rates of interaction. However, often it is not possible to observe the entire network traffic directly due to some constraints such as cost, physical limitations, or laws. By treating the unobserved network traffic as latent variables, we propose estimators for the true network traffic, including the anomalous activity, to use in testing for significant deviations from the baseline. We provide simulation results of a simple network of facilities and show that our estimators have superior performance over existing alternatives. Additionally, we establish that while a good estimate of the network traffic is necessary, perfect reconstruction is not required to effectively detect anomalous network activity. Instead it suffices to detect perturbations within the network at an aggregate or global scale. 
%
%
%
%

\section{Network Tomography Framework}

The problem of analyzing a partially observed network, termed as network tomography, was first proposed by \cite{Vardi1996}. They define each node that transmits messages to another node in the network as a source-destination (SD) pair, and they consider observing the total traffic on edges in the network.  Specific interest has developed in networks that also contain interior nodes, which do not generate or store traffic, but instead passively observe traffic. Modeling the network traffic with the Poisson distribution, which is a natural and well-known choice for the transmission of messages, \cite{Vanderbei94anem} proposes a maximum likelihood estimator (MLE) based on the expectation-maximization (EM) approach for such a network. However, their solution is not scalable and cannot incorporate observations of total traffic on interior nodes.

The existing works assume dense networks (i.e., almost all nodes communicate with each other). Thus, they are framed as under-determined linear inverse problems and do not have a unique solution. Additionally, they assume that the edge structure in the network is known and fixed, thus they cannot account for edge structure that changes over time or unknown anomalous edges. 

\begin{figure}[h] 
  \centering
  \includegraphics[width=.9\linewidth]{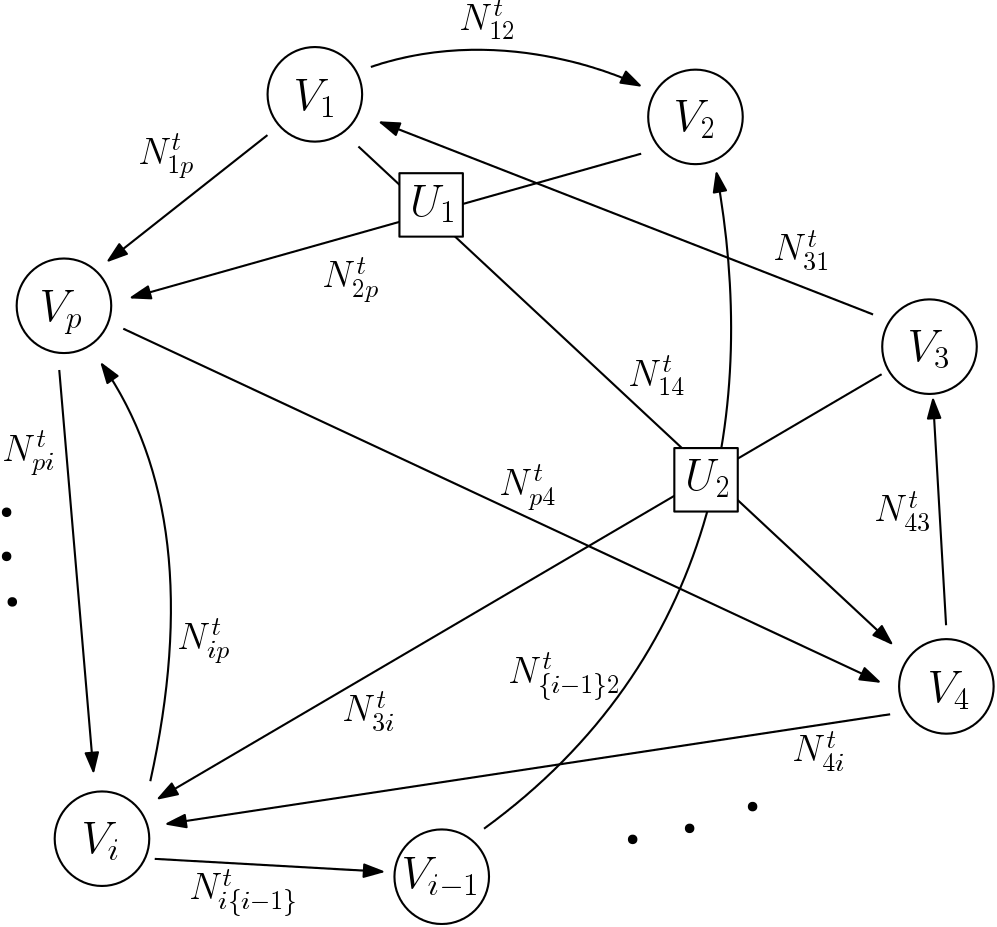}
  \caption{Proposed Network: $V_i$ - exterior nodes, $U_i$ - interior nodes, $N^t_{ij}$ - messages from node $i$ to node $j$ at time point $t$}
%
%
%
%
%
%
%
%
  \label{fig:complete}
\end{figure}

We are interested in sparse networks, where it is reasonable that an exterior node (a node in an SD pair) only transmits and receives messages from a few other nodes. Because most nodes do not communicate with most of the other nodes, the network does not have a large number of edges and is considered sparse. Additionally, we do not have any assumptions constraining the edge structure in the network, which makes the problem non-trivial, but allows for the more realistic scenario of uncertain network structure where there may be anomalous edges. Using observations of the network at multiple time points, we can estimate the rates of traffic in the network, and once we have an accurate estimate of the rates, we can test for anomalous network activity by comparing it to a baseline. We give a simple diagram of a possible network that fits in our proposed framework in Figure \ref{fig:complete}. An exterior node, $V_i$, sends messages, $N^t_{ij}$, at a rate, $\Lambda_{ij}$, to another exterior node, $V_j$, at each time point, $t$. Messages can flow through interior points, such as $U_1$, but they remain unchanged. 

\begin{figure}[h] 
  \centering
  \includegraphics[width=.9\linewidth]{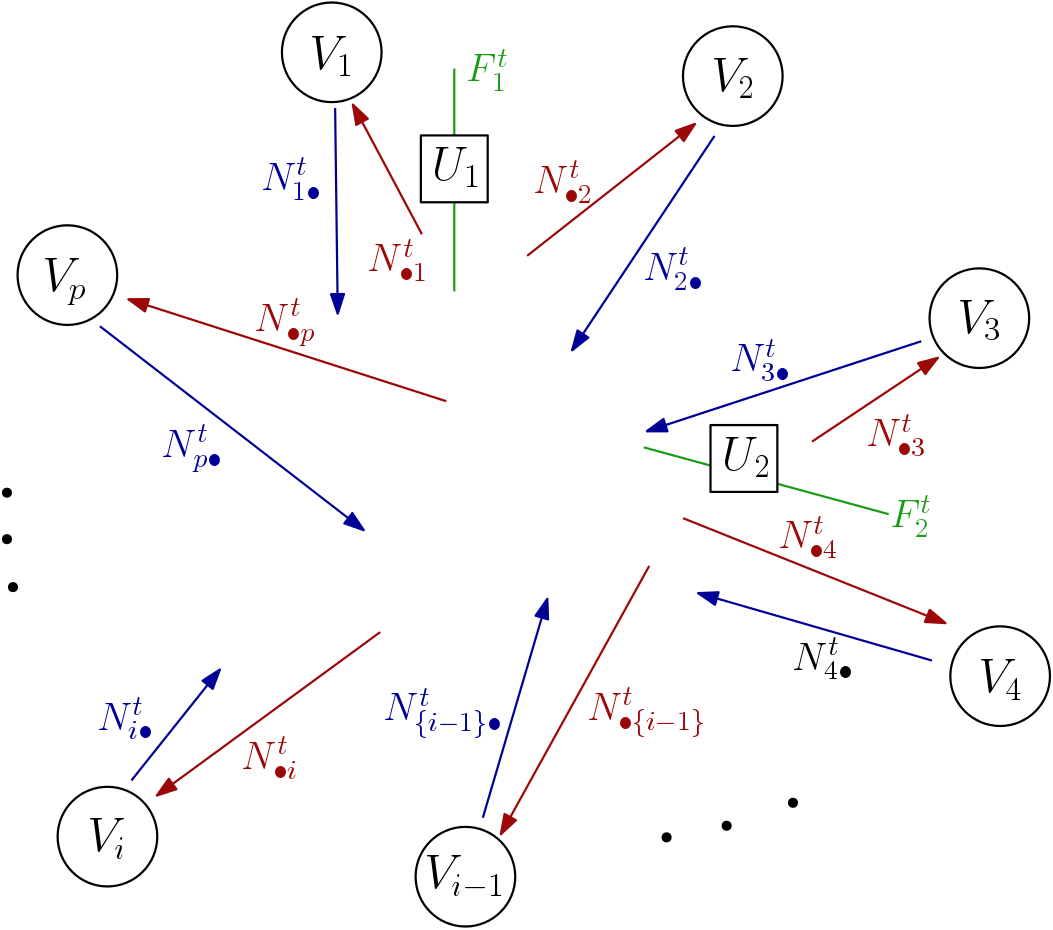}
  \caption{Actual Observed Network: $N^t_{i\cdot}$ - total egress of exterior nodes, $N^t_{\cdot i}$ - total ingress of exterior nodes, $F^t_i$ - total flow through interior nodes}
  \label{fig:actual}
\end{figure}

When the network is observed directly, the edge structure and rates are easily estimated with observations of the network at multiple time points. The solution is simply the Poisson maximum likelihood estimator (MLE), which has theoretical statistical guarantees. However, this is a very strong and unrealistic assumption because it would require being able to track every single message being passed in the network. Thus, we are interested in the much weaker assumption that we can only monitor the nodes themselves. Figure \ref{fig:actual} shows what we can actually observe from the network under this weaker assumption.

Since we can only monitor the nodes, we can only observe the total ingress and egress of the exterior nodes. Thus we know an exterior node, $V_i$, transmits $N^t_{i\cdot}$ messages and receives $N^t_{\cdot i}$ messages, but we do not know which of the other nodes it is interacting with. We can also observe the flow through interior nodes, but we cannot distinguish where the messages come from or are going to. For instance, in Figure \ref{fig:complete}, an interior node, such as $U_1$, will observe all messages, $F_1^t = \{ N^t_{14}, N^t_{2P} \}$, that flow through it, but it will not be able to distinguish the number of messages from each SD pair or if all the SD pairs actually send messages.

\section{Proposed Algorithms}

Now that we have established a framework to define the characteristics of the network, we can model how the rates are distributed. We assume \textit{a priori} that the distribution of the rates is centered around some baseline rates $\Lambda_{0 \, ij}$, which are the expected rates when there is no anomalous activity. We then update this prior distribution using the observations $\mathcal{D}$ (the total ingress, egress, and flow) in order to get a posterior distribution of the rates $ \text{P}(\Lambda | \mathcal{D}) $, which does account for potential anomalous activity. By modeling the posterior distribution of the rates, we can get statistical estimators, such as the posterior mode, and test if they differ significantly from the baseline rates. Below we propose two models for estimating the mode of two different posterior distributions. 

\subsection{Poisson Hierarchical Model}

\begin{figure}[h] 
  \centering
  \includegraphics[width=\linewidth]{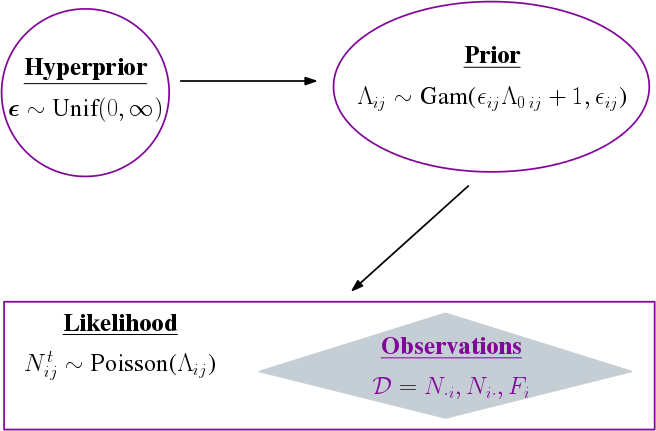}
  \caption{The statistical process believed to underlie our network.}
  \label{fig:gen_model}
  
%
%
%
%
%
%
\end{figure}

Our first model is a generative model that assumes a series of statistical distributions govern the generation of the network. This generative process is shown in Figure \ref{fig:gen_model}. 

The model assumes that \textit{a priori}, the rates of traffic in the network, $\Lambda_{ij}$, come from gamma distributions with modes at $\Lambda_{0 \, ij}$. The prior variance is controlled by the hyperparameters $\epsilon_{ij}$, which represent our belief in these baseline rates. So as $ \epsilon \rightarrow 0 $, the prior variance $ \rightarrow \infty $, and the prior becomes non-informative because we have no confidence in the baseline, while as $ \epsilon \rightarrow \infty $, the prior variance $ \rightarrow 0 $, and the prior becomes the point $ \Lambda_{0 \, ij}$ because we are certain the baseline is correct. We put a positive Uniform distribution as the hyperprior on $\epsilon_{ij}$ to indicate it is random and can take any real positive number.

If we could directly observe the network, it is natural to model the likelihood for each SD pair, $ij$, as a Poisson distribution because the rates, $\Lambda_{ij}$, generate data in the form of a discrete number of hits (messages), $N^t_{ij}$, at every time point $t$. However, since we cannot directly observe the network, we must first estimate the traffic, $N^t_{ij}$, from what we can observe, $\mathcal{D}$ (the total ingress, egress, and flows). This procedure of iteratively estimating the complete network and using this estimate to optimize for the posterior mode is done using the EM algorithm \cite{dempster1977maximum}. We show this explicitly for our model in Algorithm \ref{alg:EM}. 

\begin{algorithm}[h]
\caption{Hierarchical Poisson with EM}
\textbf{initialize:} baseline rates $\Lambda_{0 \, ij}$, $\hat{\Lambda}^{0}_{ij} = $ initialization points 
\begin{algorithmic}
\Repeat \, $ k = 1 \ldots $
\EStep
\State Estimate the network edges, $ \hat{N}_{ij} $, by its expected value \State given observations, $\mathcal{D}$, and previous M-Step's rate \State estimates, $\hat{\Lambda}^{k-1}_{ij}$
\MStep
\State Optimize for new belief estimators, $\hat{\epsilon}^k_{ij}$
\State Optimize for new rate estimators, $\hat{\Lambda}^{k}_{ij}$, from the mode \State of $\text{P}(\Lambda | \mathcal{D}, \hat{N}_{ij}, \hat{\epsilon}^k_{ij}) $, the posterior distribution with the \State E-step's estimated network edges
\Until convergence \\
\Return $ \hat{\Lambda}_{ij}^{k+1} $
\end{algorithmic}
\label{alg:EM}
\end{algorithm}

The weakness with this model is that often our posterior has a lot of local maxima and the EM algorithm is sensitive to the initialization point because it will converge to the closest maximum. Thus, it requires many random restarts or a good starting point to converge to a good local maximum. Additionally, if the starting point is not very good, the EM algorithm can take a long time to converge.

\subsection{Minimum Relative Entropy Model}
\begin{figure}[h] 
  \centering
  \includegraphics[width=\linewidth]{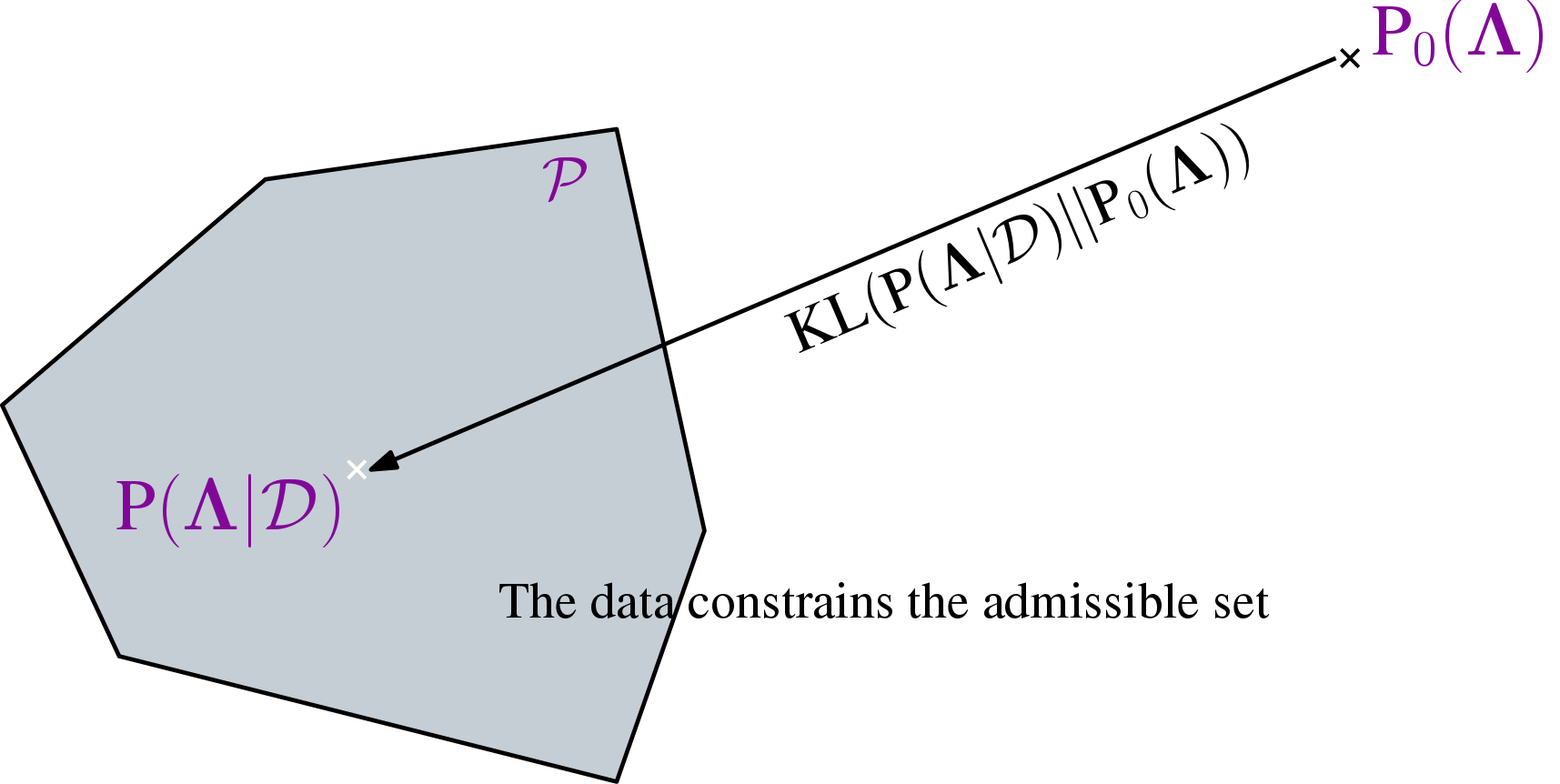}
  \caption{A projection of the prior, $P_0(\Lambda)$, onto a feasible set $\mathcal{P}$ of posterior distributions that satisfy the observed data, $\mathcal{D}$.}
  \label{fig:MRE}
\end{figure}

Instead of modeling the entire statistical process that generates our data, we can instead use the principle of minimum relative entropy (MRE) \cite{cover2012elements}. Geometrically, this is simply an information projection of the prior distribution, as shown in Figure \ref{fig:MRE}. 

The minimum relative entropy distribution is the posterior that is closest to a given prior distribution and lies in a feasible set, $\mathcal{P}$. This feasible set is formed from constraints that require the moments of the posterior to match properties of the observations, $\mathcal{D}$ (the total ingress, egress, and flows). Closeness is measured by the Kullback-Leibler (KL) divergence, which is a standard measure of difference between probability distributions. It is reasonable to want the closest posterior distribution because anomalous activity is rare, so the distribution of the actual rates, $\Lambda_{ij}$, should be similar to the distribution of the baselines rates $\Lambda_{0 \, ij}$.

When the chosen priors are independent Laplace distributions with mean $ \Lambda_{0 \, ij} $ and scale $1$, the mode of this MRE distribution is actually a relaxation of the solution to the constrained optimization problem: $ \underset{\bm{\Lambda} \in \mathbb{R}^{+}}{\arg\min} || \bm{\Lambda} - \bm{\Lambda}_0 ||_1 $ subject to the data equality constraints. This constrained optimization problem can be solved efficiently with interior point methods. 

Because the MRE distribution does not have distributional assumptions on the likelihood, 
%
%
%
%
%
%
its update of the prior distribution to a posterior is more relaxed. This allows the posterior to be more robust to the model mismatch problems; however, it also means the posterior is comparatively not as accurate as a specific likelihood assumption (like the hierarchical Poisson one) when the model matches the observations well.

\section{Simulation Results}

We simulate networks with 10 facilities (nodes) where each facility has a 65\% chance of transmitting messages to any other facility. We draw the baseline rates $L_{0 \, ij}$ from Gamma(1.75, 1) distributions and the diversion rates from Gamma(0.75, 1) distributions. The probability of a diversion between any two facilities is 20\%. All our simulations are averaged over 200 trials. Our simulations are just proof of concepts, so we let them be as general as possible to show our models are not restricted to specific scenarios. Our models can easily incorporate a more realistic fuel cycle simulation with specific types of facilities and facility interactions.

\begin{figure}[h] 
  \centering
  \includegraphics[width=.95\linewidth]{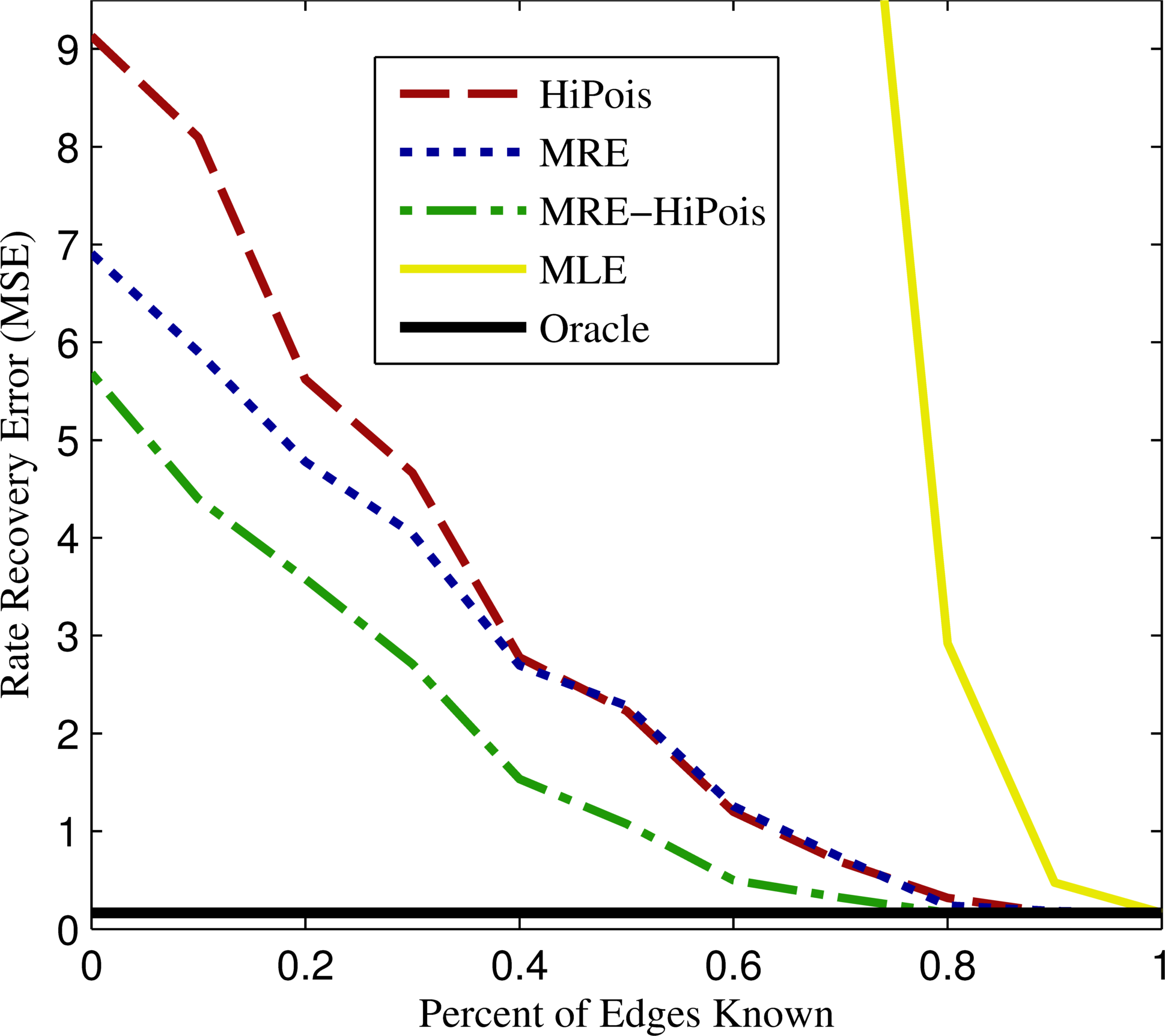}
  \caption{The error as the number of edges observed in the network increases. Note the Oracle is the theoretical lower bound.}
  \label{fig:MSE}
\end{figure}

Figure \ref{fig:MSE} shows the error of the estimated rate matrices when compared to the true rate matrix (mean squared error). Here, we also consider the case in which a percentage of the edges are also observed, in addition to the facilities. As the percentage of the edges observed in the network increases, 
%
%
%
%
all estimators' errors decrease to the Oracle estimator's error. The Oracle estimator achieves the Cramer-Rao lower bound, so it is used to show the lowest error possible; however it is not a realistic estimator because it requires observing the network directly. When the percentage of edges observed is smaller, the Hierarchical Poisson estimator (red) does worse than the MRE estimator (blue) because the EM algorithm in the former model is more sensitive to the initialization point. However, when we initialize the EM algorithm of the Hierarchical Poisson Model at the MRE estimator (green), the final estimator is significantly better everywhere. The Poisson MLE (yellow), which is a scalable version of the model in \cite{Vanderbei94anem}, is currently the standard for dealing with estimating rates of partially observed Poisson networks, but unless almost all of the network is observed, it does extremely poor. 

\begin{figure}[h] 
  \centering
  \includegraphics[width=\linewidth]{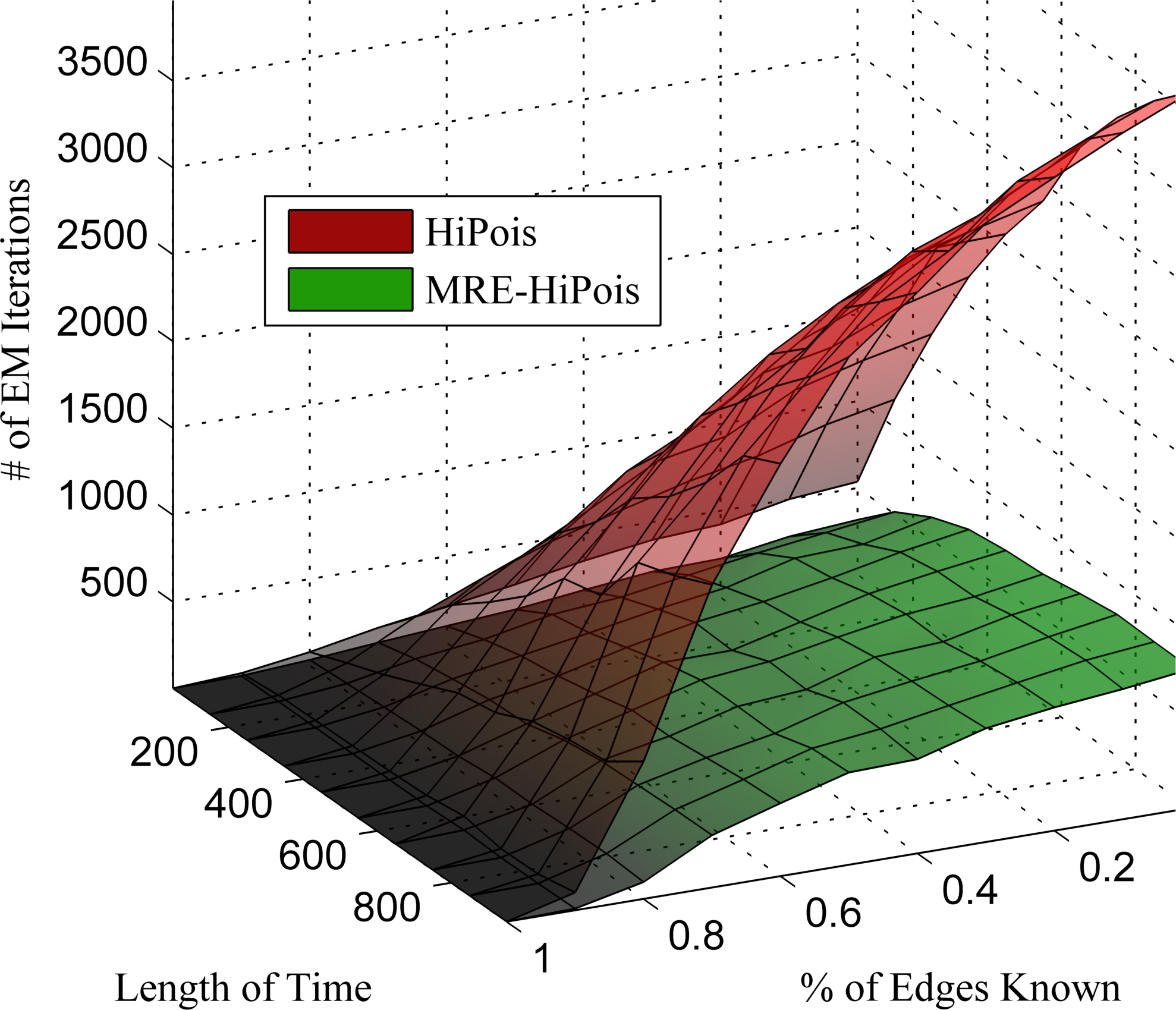}
  \caption{The number of iterations required for the EM algorithm to converge as the observation time and number of edges observed vary.}
  \label{fig:iter}
\end{figure}

Additionally, Figure \ref{fig:iter} shows that the EM algorithm in the standard Hierarchical Poisson Model (random initialization points) takes a long time to converge; especially when the observation time is large 
%
%
%
%
%
%
%
%
and the number of observed edges is small. This is because, in this regime, the likelihood becomes extremely complicated, and the EM algorithm has a difficult time deciding on the best of the nearby local maxima. But, by initializing at the MRE estimator, the EM algorithm converges much faster because it is already, on average, much closer to a much better maximum.

\begin{figure}[H] 
  \centering
  \includegraphics[width=.95\linewidth]{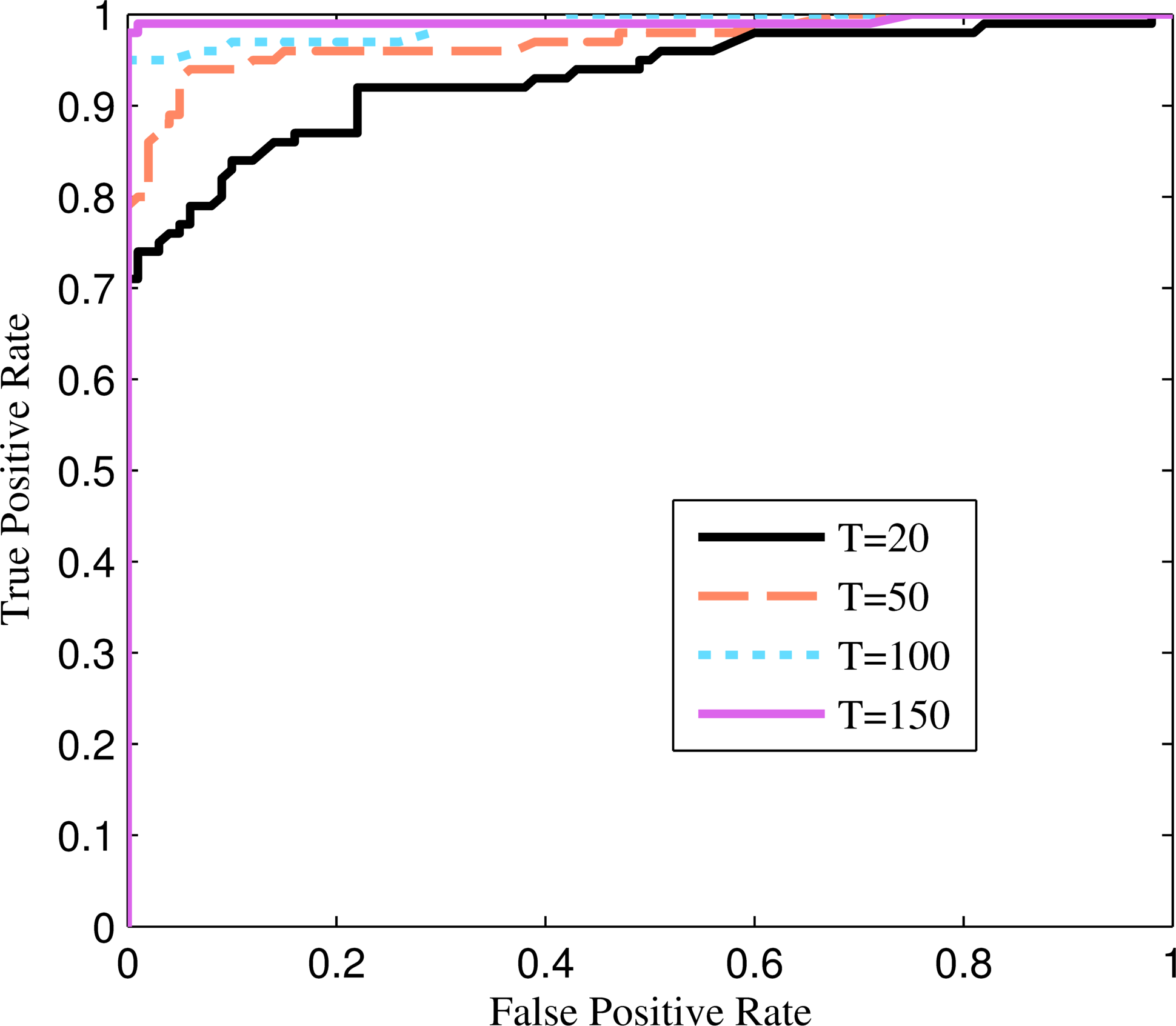}
  \caption{ROC curves of the estimator from the combined model (MRE-HiPois) for various time intervals, $T$, where none of edges in the network are observed. 50\% of the 200 trials have diversions.}
  \label{fig:ROC}
\end{figure}

Once we have estimators, $\hat{\Lambda}$, for the true rates, we can test for anomalies by seeing how far they diverge from the baseline rates. This can be done by declaring that at least one anomaly has occurred in the given observation time, $T$, if $|| \hat{\Lambda} - \Lambda_0 ||_F > \tau$, where the threshold $\tau$ is set to control the false positives. Figure \ref{fig:ROC} shows that as we increase our observation time, the ability to detect a diversion increases for any given false positive rate. With an observation time of only 150 time points, our model can detect diversions almost perfectly.

\begin{figure}[H] 
  \centering
  \includegraphics[width=.9\linewidth]{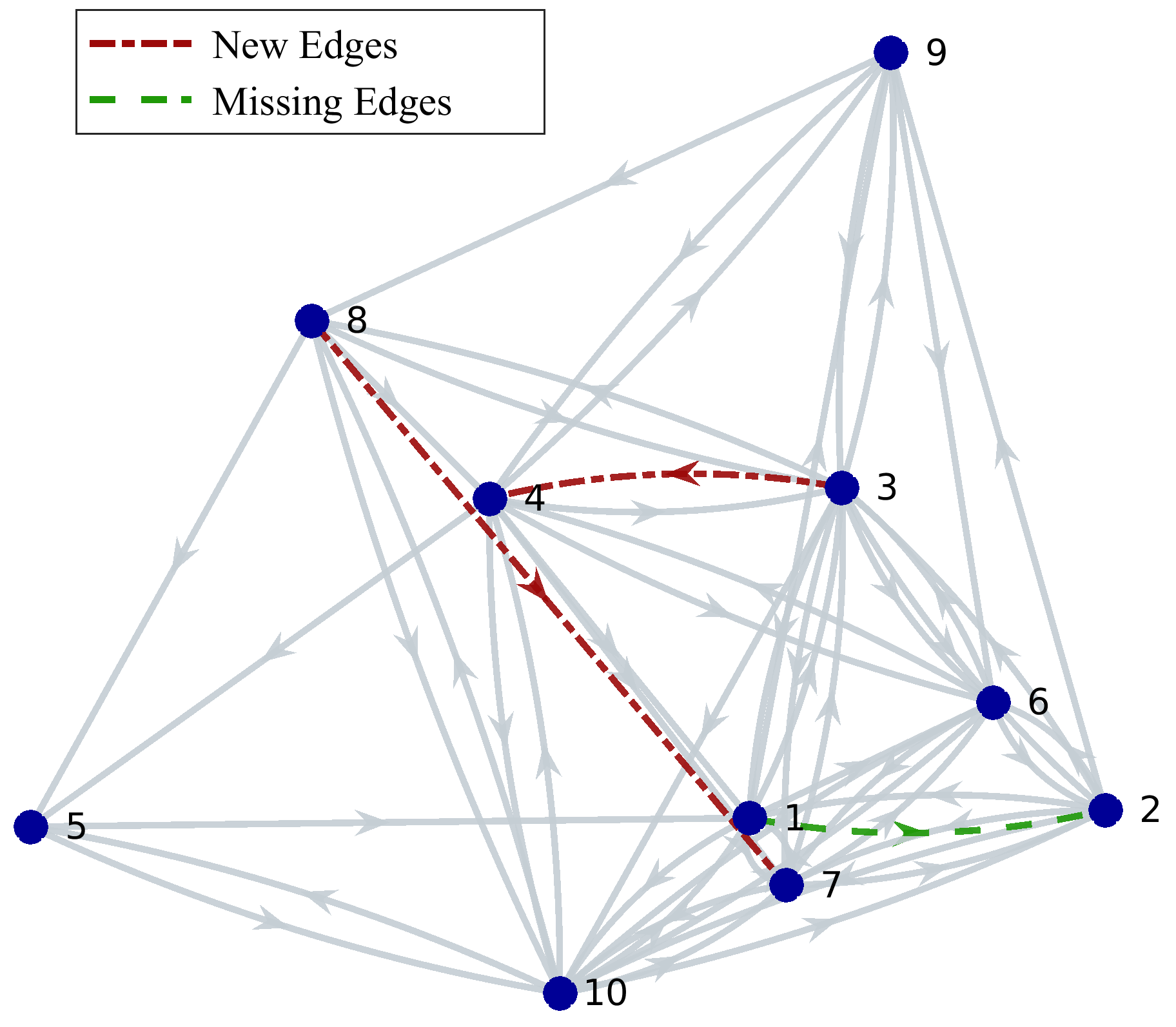}
  \caption{The estimated network using the combined model (MRE-HiPois). The model is able to correctly identify two new edges and one missing edge compared to the \textit{a priori} assumption.}
  \label{fig:anomaly}
  
%
%
%
%

\end{figure}

We show one instance of a network from our simulation in Figure \ref{fig:anomaly}. Our estimator from the combined model (MRE-HiPois) is able to detect diversions both in the form of new edges (anomalous network traffic where there should not be any) and in the form of missing edges (\textit{a priori} assumed network traffic, but found to be missing). 

\section{Conclusion}

We have developed a framework and models for estimating the traffic rates in partially observed networks. Our framework is realistic and robust in that, at minimum, it only requires observing the total egress and ingress of the nodes and it has no fixed assumptions of edge structure. In terms of a nuclear fuel cycle network, our framework necessitates only monitoring the facilities instead of all transportation of materials within the network. And, since it does not fix the edge structure to a given design, our framework allows our estimators to handle noisy observations and anomalous activity, such as diversions. Through simulations, we show that our models are scalable and extremely accurate, both in estimating the rates of the network traffic and in detecting diversions. 

\section{Acknowledgments}
\noindent This work was funded by the Consortium for Verification Technology under Department of Energy National Nuclear Security Administration award number DE-NA0002534.


\bibliographystyle{ans}
\bibliography{bibliography}

\end{document}